\documentclass[aps,prb,twocolumn,superscriptaddress,floatfix,showpacs]{revtex4}

\usepackage{graphicx}
\usepackage{bbm}
\date{\today}

\newcommand{\Csixty}{C$_{60}$}
\newcommand{\diff}{\mathrm{d}}
\newcommand{\homo}{\mathrm{HOMO}}
\newcommand{\lumo}{\mathrm{LUMO}}
\newcommand{\img}{\mathrm{i}}

\newcommand{\ket}[1]{| #1 \rangle}
\newcommand{\bracket}[2]{\langle #1 | #2 \rangle}
\newcommand{\efermi}{E_\mathrm{F}}
\newcommand{\mat}[1]{\mathbf{#1}}
\newcommand{\sub}[1]{\mathrm{#1}}
\newcommand{\gmat}[1]{\mathbf{\tilde{#1}}}
\newcommand{\cc}{\relax}

\newcommand{\myref}[1]{(\ref{#1})}

\newcommand{\eref}[1]{Eq.\,(\ref{#1})}
\newcommand{\Fref}[1]{Figure\,\ref{#1}}
\newcommand{\fref}[1]{Fig.\,\ref{#1}}
\newcommand{\sref}[1]{Sec.\,\ref{#1}}

\begin{document}

\title{Energy resolved STM mapping of C$_{60}$ on metal surfaces: A
theoretical study}


\author{Mario De Menech}
\affiliation{Max--Planck--Institut f\"ur Physik komplexer Systeme\\
  N\"othnitzer Str.\,38, 01187 Dresden, Germany}
\affiliation{Theoretische Physik, Fachbereich Naturwissenschaften, 
  Universit{\"a}t Kassel\\ Heinrich-Plett-Str.\,40, 34132 Kassel, Germany}
\author{ Ulf Saalmann}
\affiliation{Max--Planck--Institut f\"ur Physik komplexer Systeme\\
  N\"othnitzer Str.\,38, 01187 Dresden, Germany}
\author{Martin E.\ Garcia}
\affiliation{Theoretische Physik, Fachbereich Naturwissenschaften, 
  Universit{\"a}t Kassel and Center for 
Interdisciplinary Nanostructure Science and Technology (CINSaT)\\ Heinrich-Plett-Str.\,40, 34132 Kassel, Germany}
\affiliation{}

\begin{abstract}
  We present a detailed theoretical study of scanning tunneling imaging and
  spectroscopy of \Csixty\ on silver and gold surfaces, motivated by the
  recent experiments and discussion by X.\ Lu et al.\ [PRL
  \textbf{90},\,096802\,(2003) and PRB \textbf{70},\,115418\,(2004)].  The
  surface/sample/tip system is described within a self--consistent DFT based
  tight--binding model.  The topographic and conductance images are computed
  at constant current from a full self--consistent transport theory based on
  nonequilibrium Green's functions and compared with those simulated from the
  local density of states.  The molecular orbitals of \Csixty\ are clearly
  identified in the energy resolved maps, in close correspondence with the
  experimental results.  We show how the tip structure and orientation can
  affect the images. In particular, we consider the effects of truncated tips
  on the energy resolved maps.
\end{abstract}

%
%

\maketitle

\section{Introduction}

Scanning tunneling microscopy (STM) was introduced to image surface structures
in real space~\cite{BinningPRL1982}, and is nowadays used to obtain more
complete information on the local electronic properties with the scanning
tunneling spectroscopy (STS) technique.  Its operation is based on the
measurement of the tiny tunneling current flowing from a very sharp conducting
tip to the surface when a bias voltage is applied. Images are obtained by
recording the vertical displacement of the tip as it is moved across the
surface, while keeping the tunneling current constant with a precise feedback
mechanism. If the current increases due to a protrusion on the surface, the
tip is lifted to bring the current back to the set value, therefore revealing
the topography of the substrate.

Tersoff and Hamann~\cite{Tersoff1985} have shown how the images do not only
reflect the geometric structure of the surface, but also depend on the
electronic density of states of the sample, which is identified, to a first
approximation, with the differential conductance.  This observation has opened
the way to spectroscopic measurements with a spatial resolution which is by
far not accessible by other surface science techniques. This capability is of
particular interest in the study of adsorbates at surfaces, which can be
addressed individually by the probe. The STS signal, on the other hand, is
strongly influenced by the nature of the contact and the structure of the tip,
and the extraction of information on the physical and chemical properties of
the measured sample is possible only with the support of electronic structure
calculations with an atomistic description of the
system~\cite{Corbel1999,Drakova2001,Blanco2004}.

 In this paper we discuss from a theoretical point of view the powerful
combination between microscopy and spectroscopy which is peculiar of STM/STS.
The accuracy of recent STM experiments on \Csixty\ monomers on metallic
surfaces~\cite{Lu2003,Lu2004} represents a rigorous benchmark for any
theoretical modeling of STM imaging and spectroscopy.  Quite interestingly,
energy resolved mapping of the molecular orbitals of adsorbed C$_{60}$
monomers has already been used as a control reference for even more
sophisticated experiments~\cite{Grobis2005}; similarly, we would like to
discuss in detail STM/STS mapping of this system based on Green's function
methods, in view of the investigation of adsorbed molecules and
clusters~\cite{DeMenech2005} which have not yet been experimentally
characterized with this technique. Topographic and energy resolved maps are
computed with a realistic model of the STM probe, with the simulation of the
tip trajectory over constant current surfaces; the effects of the tip
structure on the rendered images are illustrated in detail, showing that
truncated tips may duplicate the characteristic features depending on the
orientation of the topmost atoms with respect to the sample.  Also, we refine
previous calculations based solely on the local density of states (LDOS) of
the supported molecule~\cite{Lu2003,Lu2004}.  We confirm that the images
obtained from the LDOS can effectively approximate the patterns observed in
the STS maps, with the additional piece of advice that the simulated probe
trajectory should be taken at the correct distance from the molecule.

The paper is organized as follows.  The theoretical framework is briefly
reviewed in \sref{sec:theory}. The experienced reader may skip this part and
move directly to the results in \sref{sec:results}.  Equilibrium electronic
properties of the supported \Csixty\ are discussed in \sref{sec:total_DOS},
while the calculations based on transport theory are in \sref{sec:transport},
extended by simulation of topographic and spectroscopic images with different
types of tips in \sref{sec:imaging}.  Finally, in \sref{sec:LDOS} we specify
how STS maps can be modeled using the local density of states.

\section{Theory}
\label{sec:theory}

We consider a fully atomistic description of surface, molecule and STM probe.
 The geometry of substrate and the STM electrode are fixed from bulk
 properties, i.e. surface relaxation effects are neglected, and the structure
 of the molecule is assumed to remain unchanged after adsorption.  The
 transport properties are computed with a microscopic approach using
 non--equilibrium Green's functions, and the electronic structure is modeled
 within a single--particle picture, based on a self--consistent tight--binding
 (TB) model which is parameterized from density functional theory (DFT).

\subsection{TB--DFT model}
\label{sec:TB-DFT}

Electron wavefunctions are approximated by an expansion in an atomic--like
orbital basis~\cite{Porezag1995}, which is determined by solving the
Kohn--Sham equations of the individual atoms within the local density
approximation (LDA). The TB Hamilton matrix expressed in the localized basis
consists only of two--center integrals which can be obtained exactly by using
a Slater type functions for the atomic orbitals. This scheme is refined to
include charge self--consistency by expanding the DFT energy functional to the
second order in the charge--density
fluctuations~\cite{SaalmannThesis,Elstner1998}. Such an expansion leads to the
appearance of an additional term to complete the TB Hamiltonian that contains
the contribution of the net atomic charges on the local effective potential.
Like in a full DFT calculation, self--consistency is obtained with iterative
methods, such that the local density, given by the sum of the (approximated)
Kohn--Sham eigenfunctions, is consistent with the effective potential. The
main advantage of this approach lays in its simplicity and computational
efficiency, while basic key effects like charge transfer and redistribution
are still handled correctly.

\subsection{Equilibrium electronic properties of adsorbates}
\label{sec:equilibrium}

Inglesfield~\cite{Inglesfield1981b,Inglesfield1981a} introduced the concept of
embedding potential to perform real space electronic structure calculations in
a domain surrounding impurities or defects in an otherwise perfect
crystal. The form of the embedding potential is derived from the Green's
function of the solid, and is included in an effective Schr\"odinger equation,
to be solved only in the region of interest, ensuring the correct matching of
the solutions with the wavefunction in the rest of the solid. A complementary
approach considers the defect as a perturbation of the ideal system, and has
been first proposed by Williams and Lang~\cite{Williams1982}. A localized
atomic orbital basis $\left\{\ket{\varphi_\mu(r)}\right\}$ is introduced to
represent in matrix form the Hamiltonian $\gmat{H}$ and the (retarded) Green's
function $\gmat{G}$, defined by
\begin{equation}
\left[(E{+}\img\eta)\:\gmat{S}  -\gmat{H} \right] \gmat{G} (E) 
= \mat{\mathbbm{1}},
\hskip 0.2 cm \eta\to 0^+,
\label{eq:green_def}
\end{equation}
$\gmat{S}\equiv S_{\mu\nu}=\bracket{\varphi_\mu}{\varphi_\nu}$ being the
overlap matrix.  The indices for matrices with a tilde run over the atomic
orbitals of the surface and the cluster.

Measurable quantities can be computed directly from the Green's function; for
example, from the spectral density matrix
\begin{equation}
A_{\mu\nu}(E)=-\frac{2}{\pi}  \mathrm{Im} \left[G_{\mu\nu}(E)\right]
\label{eq:spectral_density_matrix}
\end{equation}
(the factor 2 is explicity included to point out that we consider a
spin--unpolarized system), it is possible to compute the local density of
states
\begin{equation}
\rho(\mathbf{r},E)=\sum_{\mu\nu} A_{\mu\nu}(E)
\varphi_\mu(\mathbf{r}) \varphi_\nu(\mathbf{r})
\label{eq:density_matrix_eq}
\end{equation}
and therefore the density
\begin{equation}
n_\mathrm{eq}(\mathbf{r})=\int_{-\infty}^{\efermi}\mathrm{d}E\;
\rho(\mathbf{r},E); 
\label{eq:density}
\end{equation}
where $\efermi$ is the Fermi level (degenerate case). 

The localized basis allows to translate the cluster--surface partitioning of
the system into the block structure for the overlap, the Hamilton, and the
Green's function matrices of~\eref{eq:green_def}.  The Green's function of the
isolated cluster $\mat{G}^0\cc$ is modified by the presence of the surface,
and can be obtained from the solution of the Dyson equation
\begin{equation}
\mat{G}\cc=\mat{G}^0\cc+\mat{G}^0\cc \mat{\Sigma}\cc  
\mat{G}\cc
\label{eq:Dyson}
\end{equation}
where $\mat{G}\cc$ refers to the Green's function of the ``embedded'' cluster,
and $\mat{\Sigma}\cc$ is the matrix analogue~\cite{Baraff1986} of
Inglesfield's embedding potential.  Note that, in contrast to
\eref{eq:green_def}, all matrices in \eref{eq:Dyson} are defined for orbitals
at the cluster only.  $\mat{\Sigma}\cc$ can be expressed as a self--energy
\begin{equation}
\mat{\Sigma}\cc (E)= 
\mat{V}(E)
\mat{G}^0_\sub{s}(E)
\mat{V}^\dagger(E),
\label{eq:self-energy}
\end{equation}
thereby including the scattering processes due to the interaction with the
surface, given by the combination of the unperturbed surface Green's function
$\mat{G}^0_\sub{s}$ and the tunneling matrix $\mat{V}=E\:\mat{S}_\sub{cs}
-\mat{H}_\sub{cs}$.  $\mat{S}_\sub{cs}$ and $\mat{H}_\sub{cs}$ are the overlap
and Hamilton matrices between orbitals at the surface and the cluster.  For
an orthogonal basis $\mat{S}_\sub{cs}$ would vanish and $\mat{V}$ would be
independent of the energy $E$.

Having chosen a localized orbital basis, only a limited number of elements in
the tunneling matrix are different from zero, and $\mat{G}^0_\sub{s}$ needs to
be evaluated just for the surface atoms which are within a finite distance
from the supported cluster.  The two--dimensional periodic structure of the
surface allows to reduce the corresponding overlap and Hamilton matrices in
a block--diagonal form, based on the transformation to the two--dimensional
momentum space. The surface is then partitioned in a series of adjacent
layers, and the Green's function for the first layer, the needed submatrix of
$\mat{G}^0_\sub{s}$, can be efficiently computed for any $\mathbf{k}$--point
using recursive methods~\cite{LopezSancho1985}. The conversion to the real
space representation is finally obtained by integrating over a finite number of
$\mathbf{k}$--values in the first Brillouin zone (we sampled over a regular
grid with at least 1600 points).

Charge self--consistency requires that the effective potential entering in the
system hamiltonian leads to a charge density which generates the same
effective potential. It is obtained with an iterative scheme, by computing the
local charge densities from $\mat{G}\cc$, to be fed into the local effective
potential of the cluster Hamiltonian~\cite{Elstner1998}, therefore leading to
a new cluster Green's function, which provides the density for the next
iteration, until convergence is reached. The integral of~\eref{eq:density} is
evaluated most efficiently taking advantage of the analytic structure of the
Green's function, with a contour integration in the upper complex
plane~\cite{Williams1982,Brandbyge2002}.  Since the substrate is described as
an ideal surface, the effect of screening due to the charge redistribution in
the metal is achieved by including image charges, which generate an external
field acting on the cluster region, and provide the correct boundary
conditions for the Poisson equation, such that the electrostatic potential on
the first atomic plane of the surface is uniform and set to zero.

\begin{figure*}[t]
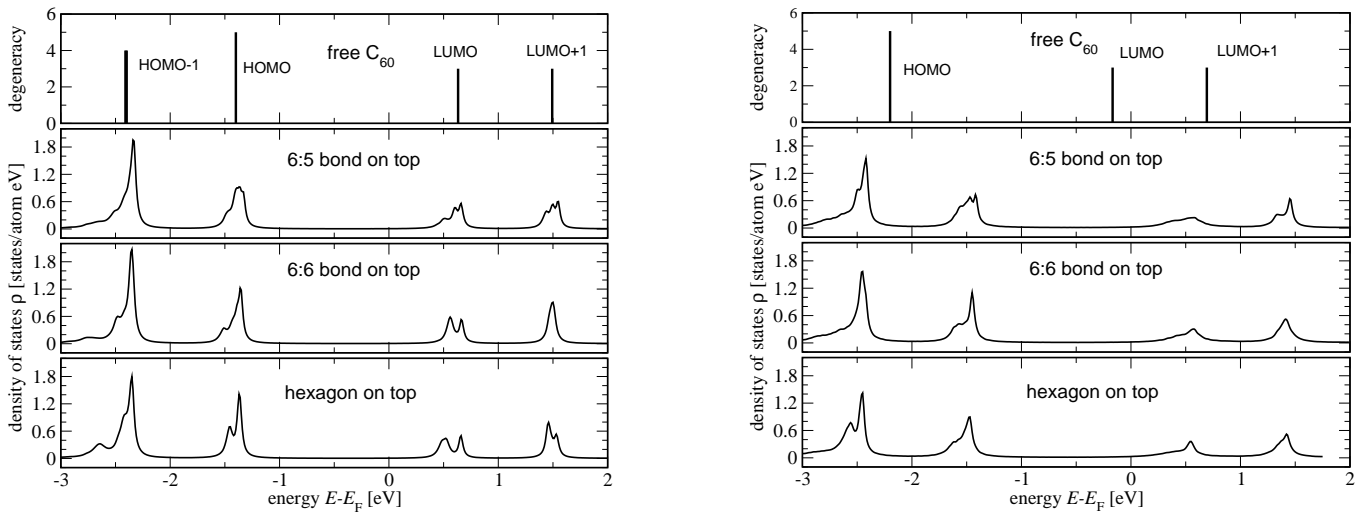

\begin{center}
\includegraphics[width=0.45\textwidth]{Fig1a_dos.eps}
\hfill
\includegraphics[width=0.45\textwidth]{Fig1b_dos.eps}
\end{center}
\caption{Total DOS $\rho(E)$ of \Csixty\ with different orientations on
  Au(111), \textsl{left column}, and Ag(111), \textsl{right column}.
  The top figure shows the position of the HOMO, LUMO and LUMO$+$1 orbitals
  for the free molecule relative to the Fermi level of the Au and the Ag
  surface, respectively.}
\label{fig:dosC60Au111Ag111}
\end{figure*}

\subsection{Transport}

 The expression for the density in \eref{eq:density} is valid only in
equilibrium conditions. When the sample is probed with an STM tip at nonzero
bias $V$, the local charge is obtained from the correlation function
$\mat{G}^<$ defined by \cite{DattaBook}
\begin{equation}
\mat{G}^<= \mat{G}\cc \mat{\Sigma}^< \mat{G}\cc^\dagger,
\end{equation}
where the lesser self--energy
\begin{equation} 
\label{eq:selftipsurf}
\mat{\Sigma}^<= \img \left[f(E{-}\mu_\sub{s}) \mat{\Gamma}^\sub{s}
+f(E{-}\mu_\sub{t})\mat{\Gamma}^\sub{t}\right],
\end{equation}
depends on the chemical potentials $\mu_\sub{s}=\efermi$
and $\mu_\sub{t}=\mu_\sub{s}+|e|V=\efermi+|e|V$ 
of the surface and the tip, respectively.   
In \eref{eq:selftipsurf},
$f$ is the Fermi function, and $\mat{\Gamma}^\sub{s}$ and 
$\mat{\Gamma}^\sub{t}$ are the broadening matrices. 
They include the interaction with the surface and
the tip through the respective self--energies
\begin{equation} 
\mat{\Gamma}^\sub{s}=
\img\left[\mat{\Sigma}^\sub{s}\cc-[\mat{\Sigma}\cc^\sub{s}]
^\dagger\right],
\hskip 0.8 cm
\mat{\Gamma}^\sub{t}=
\img\left[\mat{\Sigma}^\sub{t}\cc-[\mat{\Sigma}\cc^\sub{t}]
^\dagger\right],
\end{equation}
where the cluster--tip self--energy, $\mat{\Sigma}^\sub{t}\cc$, is defined as
for the surface case in~\eref{eq:self-energy}.  The non-equilibrium analogue
of \eref{eq:density} reads
\begin{equation}
n(\mathbf{r},V)=\frac{1}{\pi \img} \int^{\infty}_{-\infty} \diff E\;
G_{\mu\nu}^<(E,V) \varphi_\mu(\mathbf{r}) \varphi_\nu(\mathbf{r}),
\label{eq:density_matrix_neq}
\end{equation}
and enters in self--consistent relaxation of the electronic charge in the
cluster, as for the equilibrium case.  The correlation function, and therefore
the non--equilibrium density, depends directly on the applied bias $V$ due to
the shift of the tip chemical potential $\mu_\sub{t}$. The numerical procedure
for the computation of the electron density splits the integral
in~\eref{eq:density_matrix_neq} into equilibrium and non--equilibrium
contributions, exploiting the same contour integration used to evaluate the
equilibrium density~\cite{Brandbyge2002}.

In the crudest approximation, the tip electrode is assumed to be a flat metal
surface, to be treated exactly like the one supporting the sample.  The
boundary conditions for the Poisson equation are that the surfaces of the two
electrodes have a constant uniform potential differing by the bias voltage,
and the correct electrostatics is obtained using a series of image charges on
both sides. This choice of boundary conditions implies that $\mat{G}^<$
depends on the external potential generated by the two electrodes, which
enters both in the cluster and tunneling Hamiltonians, and therefore
determines the cluster Green's function, now $\mat{G}\cc(E,V)$, and the
broadening matrices. Realistic tips are easily included into the model by
considering the tip atoms as part of the cluster, which will then just be a
larger sample sandwiched between the two parallel electrodes. The arrangement
of the tip atoms was assumed to be consistent with the crystal structure of
the underlying ideal surface.

\begin{figure*}[t]
\begin{center}
  \includegraphics[width=0.9\textwidth]{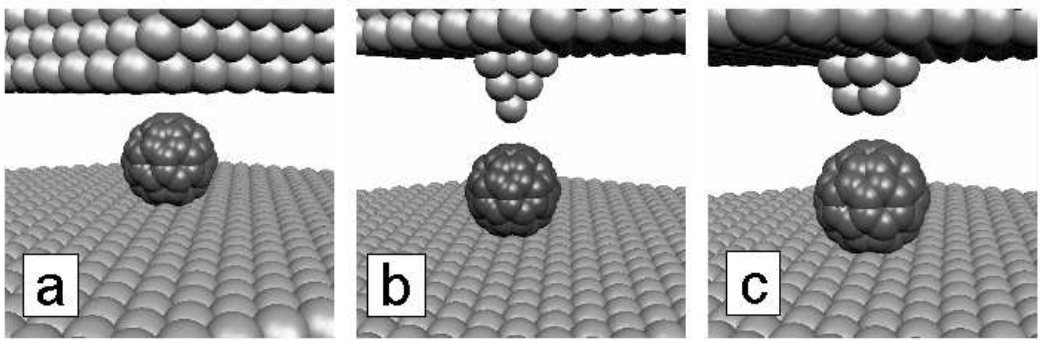}
\end{center}
\caption{Representation of the junction formed by the
  surface/\Csixty/tip/electrode. 
  a: flat electrode (no tip atom), 
  b: sharp tip (single atom at tip end), 
  c: truncated tip (three atoms at tip end).}
\label{fig:setup}
\end{figure*}
In the case of coherent transport, i.e.\ when there is no inelastic scattering
within the sample, the stationary current is given by the simple
formula~\cite{Meir1992}
\begin{equation}
I=\frac{2e}{h}\int^{\infty}_{-\infty} \diff E\; T(E,V) 
\left[f(E{-}\mu_\sub{t})-f(E{-}\mu_\sub{s})\right]
\label{eq:current}
\end{equation} 
where $T$ is the transmission function~\cite{XueNDROrganicMolecules1999}
\begin{equation}
  T(E,V)=\mathrm{Tr}\left[ \mat{G}\cc(E,V)
    \mat{\Gamma}^\sub{s}(E)\mat{G}\cc{}^\dagger(E,V)
    \mat{\Gamma}^\sub{t}(E{-}|e|V) \right].
  \label{eq:transmittance}
\end{equation}
The effects of scattering processes within the conductor due to
electron--electron or electron--phonon interactions can be treated
consistently within the non--equilibrium Green's function framework by adding
the corresponding self--energy terms~\cite{Meir1992,DattaBook}, but will not
be considered in this paper. Electron--electron interactions are incorporated
here at a mean--field level, which does not cause incoherent
effects~\cite{DattaBook}.

\section{Results}
\label{sec:results}

For the calculations presented below, the \Csixty\ was placed 
with different orientations on Ag(111) and Au(111) surfaces at
a distance of 2.4\,\AA, consistent with the value obtained from first
principles calculations for monolayers~\cite{WangChengC60monolayers2004}.
The structure of the supported molecule was not relaxed from the free
configuration.

\subsection{Total DOS of supported cluster}
\label{sec:total_DOS}

\Fref{fig:dosC60Au111Ag111} shows the total DOS of \Csixty\ for three
different orientations on Au(111) and Ag(111) surfaces.  For each of these
orientations it was observed that the number and position of the peaks remain
unchanged as the molecule is translated in the plane parallel to the surface,
showing that in both cases there is no chemical bonding with the substrate.
For \Csixty\ on Ag(111), besides the broadening and splitting of the levels of
the free molecule due to the interaction with the surface, we observe a shift
of $+0.7$\,eV associated with a net transfer of $0.3$ electrons to the
molecule, as given by the Mulliken charge analysis. In the case of the Au
substrate there is almost no charge transfer (0.01 electrons), which leaves
the position of the free cluster energy levels unchanged.
Also, we observe that the LUMO peak for the Ag case appears much more
broadened than for Au, with its lower energy tail approaching the Fermi level,
similarly to what was observed by other DFT calculations~\cite{Lu2003,Lu2004}.
The energy difference $E_\lumo-E_\homo$ is equal to 2.0\,eV in both cases,
while the experimental gap is 2.0\,eV for Ag(100) and 2.7\,eV for
Au(111)~\cite{Lu2003,Lu2004}.  The discrepancy is to be imputed mainly to the
intrinsic limits of LDA eigenvalues as a representation of electronic
excitation energies.

\begin{figure*}[t]
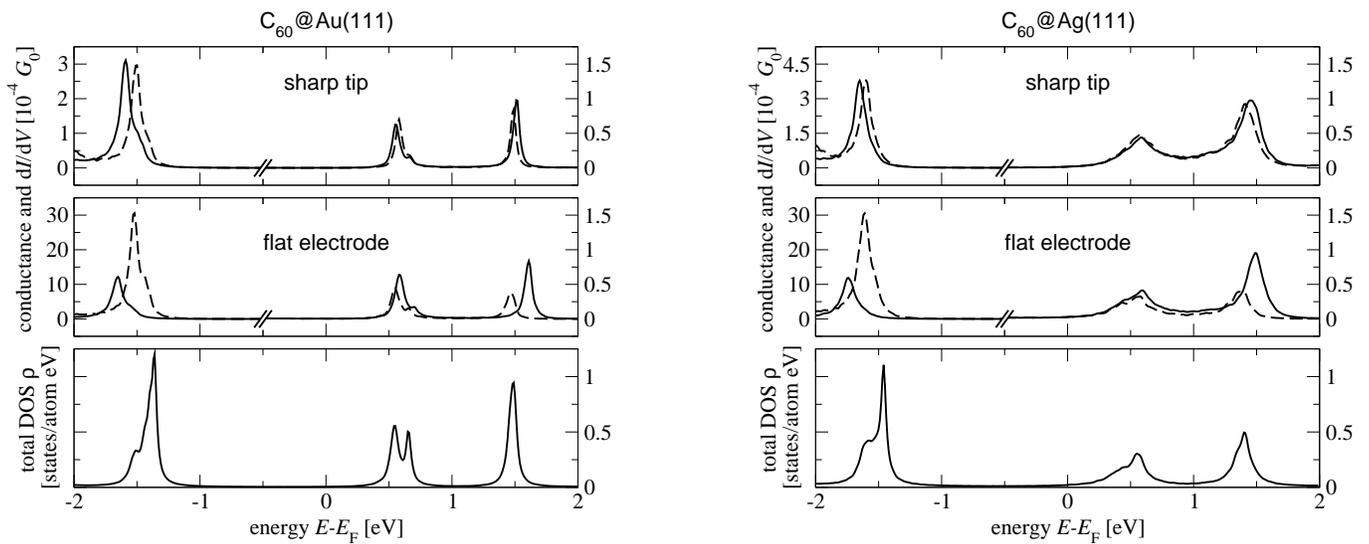

  \begin{center}
    \includegraphics[width=0.45\textwidth]{Fig3a_transport.eps}
    \hfill
    \includegraphics[width=0.45\textwidth]{Fig3b_transport.eps}
  \end{center}
  \caption{Comparison of spectral density, conductance and differential
    conductance for \Csixty\ on Au(111) and Ag(111) surfaces.  \textsl{Bottom
    panels}: Total DOS $\rho(E)$.  \textsl{Center and top panels:} Conductance
    $G$ (\textsl{dashed lines}) and differential conductance $\diff I/\diff V$
    (\textsl{solid lines}) for a probe made with the flat electrode
    (\fref{fig:setup}a) or the sharp tip (\fref{fig:setup}b), respectively,
    each placed at 5.5\,\AA\ from the top of the molecule.  Note the different
    scales for the peaks of $G$ and $\diff I/\diff V$ for $E-\efermi<-0.5$\,eV
    and $E-\efermi>-0.5$\,eV, respectively  ($G_0=2e^2/h$).  }
  \label{fig:dos_transC60}
\end{figure*}

\begin{figure*}[t]
\begin{center}
\includegraphics[height=0.75\textwidth, angle=-90]%
	{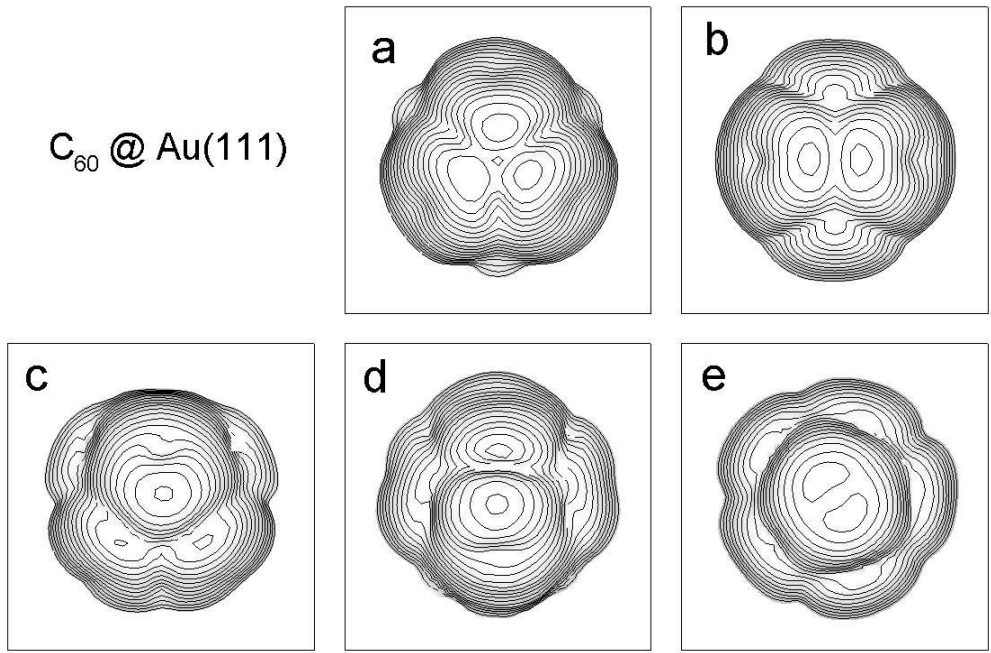}
\end{center}
\caption{Computed topographic images of \Csixty\ on Au(111) with different
  orientations; the topmost features are: (a) a hexagon ring, (b) a 6:6 bond,
  (c) a 5:6 bond, (d) an apex atom, and (e) a pentagon ring. 
  Current was set to $77.5$\,pA, the bias to $+1.8$\,V. 
  The size of each frame is $20\times20$\,\AA. 
  Compare to Fig.~1 in \textcite{Lu2004}.}
\label{fig:topography}
\end{figure*}
\begin{figure}[b]
\begin{center}
\includegraphics[width=0.45\textwidth]{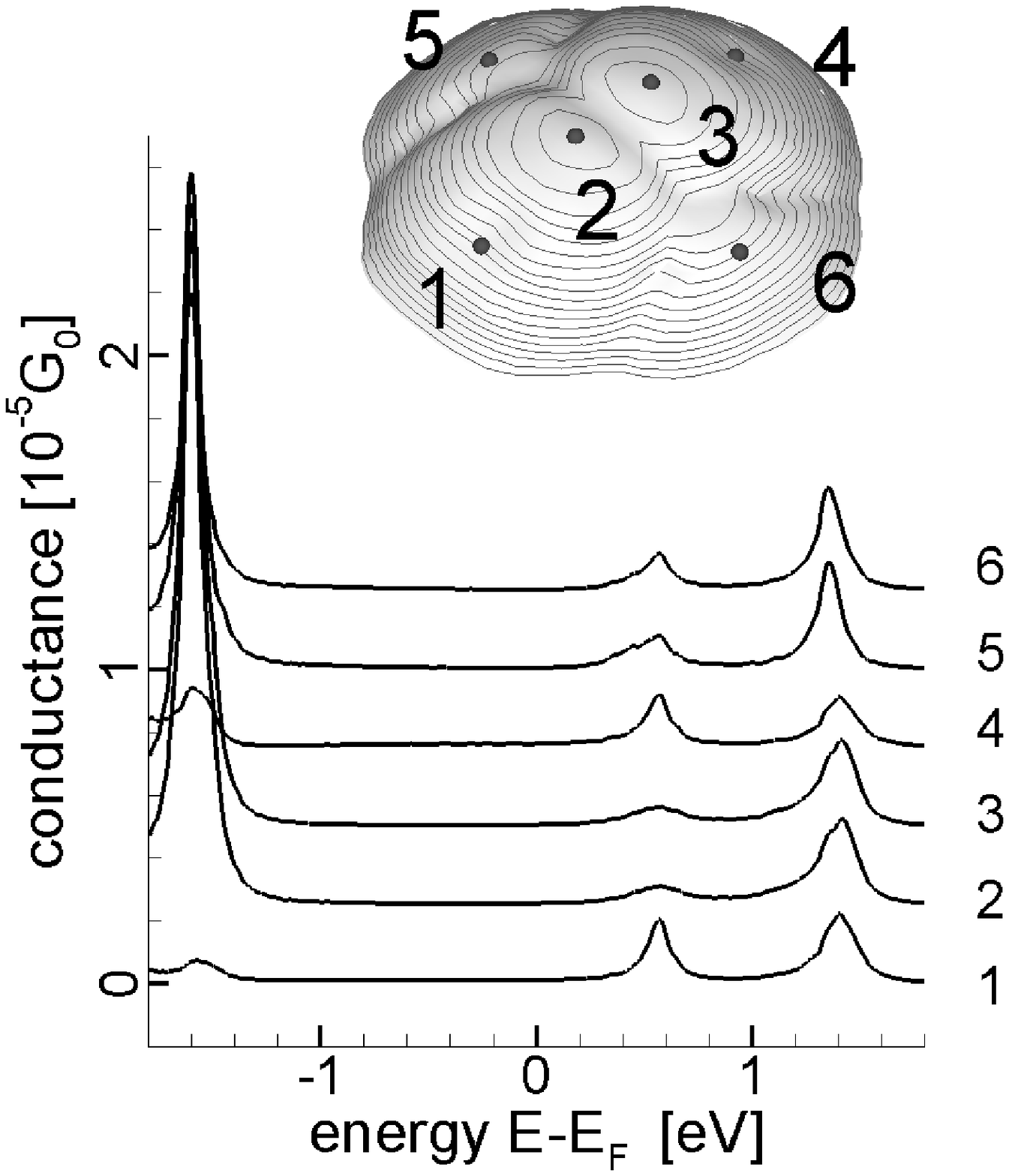}
\end{center}
\caption{Dependence of calculated STS spectra on the tip
  position for the \Csixty\ at the  Ag(111) surface with a 6:6
  bond on the top. 
  The spectra are shifted along the vertical axis for clarity.}
\label{fig:topography_spectra}
\end{figure}

\subsection{Transport calculations}
\label{sec:transport}

For the calculation of the conductance as a function of energy and of the
current--voltage characteristics we consider the combination \Csixty/tip as a
single conductor placed between two parallel, infinite, ideal surfaces
(\fref{fig:setup}).  This way of partitioning the system is convenient since
the Green's function of the surfaces can be computed efficiently (see
\sref{sec:equilibrium}).  We choose Au as element for our probe, which, in
contrast to the transition metals used in experiments, has a fairly uniform
DOS near the Fermi level, without any contribution from the d states.  This
choice allows us to single out the effects of the tip atomic structure on the
calculated maps.  We will use the term conductance for the quantity
\begin{equation}
  \label{eq:cond}
  G(E)=G_0 T(E,V{=}0), 
\end{equation}
with $G_0=2e^2/h$ and $T(E,V)$ defined in \eref{eq:transmittance}; the
transmission function is therefore  computed from the self--consistent
Hamiltonian of the system at zero bias.  In contrast, the differential
conductance $\diff I/\diff V$ is obtained from the derivative of the total
current~\myref{eq:current} for each bias value, and will be sensitive to the
external field, leading most notably to shifts of the energy levels and
therefore to the peak positions.

In~\fref{fig:dos_transC60} we compare the spectral density $\rho(E)$ of the
supported molecule with the conductance $G$ and differential conductance
$\diff I/\diff V$ curves as a function of the bias.  The latter two are
computed both for a flat electrode (\fref{fig:setup}a) and a sharp tip
consisting of ten atoms (\fref{fig:setup}b).  Conductance and total DOS
peaks appear at the same position.  However, the relative height of the peaks
changes dramatically; when switching to the sharp tip, the conduction through
the HOMO level drops by about an order of magnitude, while for the LUMO and
LUMO$+1$ there is a slight increase in the conductance.

At finite bias, the external field modifies the local charge distribution and
displaces the peaks of the differential conductance with respect to the
equilibrium energy levels.  Due to the large separation between the tip and
the \Csixty, these shifts can be appreciated only at large biases.  A similar
behavior is observed for a sharp tip, which in this case was placed centrally
above the \Csixty\ molecule.  Here, as well, the HOMO conducts current better
than the LUMO and LUMO$+1$ states; the difference in conductance, however, is
attenuated.  This behavior depends in fact on the position of the tip, and
will be discussed in the following section.

\begin{figure}[b]
\begin{center}
\includegraphics[width=0.4\textwidth]{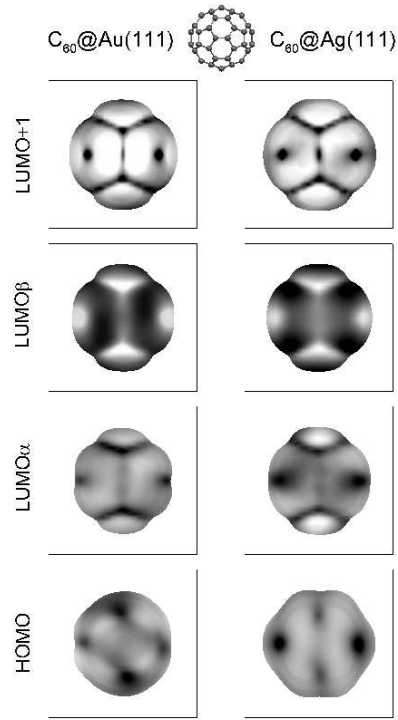}
\end{center}
\caption{Energy resolved maps of molecular orbitals for \Csixty\ on Au(111)
  and Ag(111) surfaces with 6:6 bond as the uppermost feature. The maps for
  HOMO, LUMO$\alpha$, LUMO$\beta$, and LUMO$+1$ are obtained at -1.32\,eV, 
  0.40\,eV, 0.64\,eV and 1.46\,eV for Au(111) and at -1.40\,eV, 0.30\,eV, 
  0.55\,eV and 1.36\,eV for Ag(111). The size of each frame is
  $20\times20$\,\AA.}
\label{fig:C60bond66STSmaps}
\end{figure}
\begin{figure}[b]
\begin{center}
\includegraphics[width=0.4\textwidth]{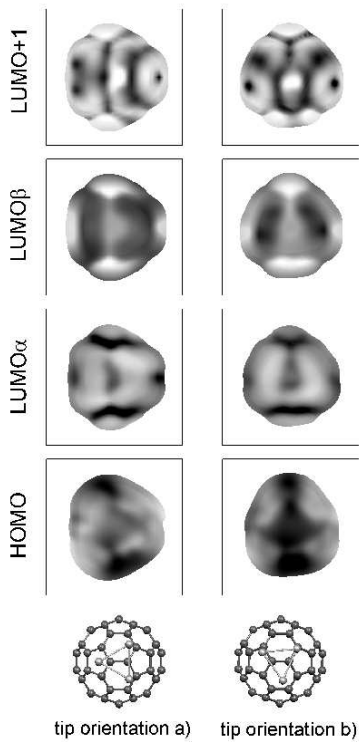}
\end{center}
\caption{Energy resolved maps with a truncated tip of \Csixty\ on Au(111) with
  the 6:6 bond on top (compare~\fref{fig:C60bond66STSmaps}).  Column a): tip
  in the reference position; column b): tip rotated by 23$^\circ$ degrees
  around the axis normal to the surface. The orientation of the three atoms at
  the bottom of the STM probe is reported for each series of images. The size
  of each frame is $20\times20$\,\AA.}
\label{fig:abnormal_images}
\end{figure}

\subsection{STM/STS imaging with realistic tip models}
\label{sec:imaging}

\subsubsection{Normal maps}

We first model STM/STS imaging with a sharp tip (\fref{fig:setup}b).  The tip
is moved on a three--dimensional finite--element mesh placed above the
molecule, with a grid spacing of 0.5\,\AA.  For each position the current $I$
and the differential conductance $\diff I/\diff V$ are computed
using~\eref{eq:current}.  The constant--current iso-surface is obtained by
interpolation, and provides the topographic image of the molecule at the given
voltage bias, as shown in~\fref{fig:topography}a--e for five different
orientations. The bias is set to $+1.8$\,V, and the pentagon rings of the
\Csixty\ bulge out in a doughnut--like shape from the spherical structure of
the molecule, allowing the direct identification of its orientation on the
surface.

Based on the topographic scan of \Csixty\ with a 6:6 bond (a carbon bond
shared by two hexagon rings) as a topmost feature reported
in~\fref{fig:topography}b, we show in \fref{fig:topography_spectra}\,
conduction curves for six tip positions.  Heights and shapes of the peaks vary
considerably as the tip is moved over the molecule, and we also observe slight
shifts in their position in the energy scale, induced by the changes in the
local electric field surrounding the STM probe.

The strong spatial dependence of the conductance allows to construct
energy--resolved maps (\fref{fig:C60bond66STSmaps}) to represent the states
associated with each peak.  The simulated STS maps are rendered by displaying
the computed $\diff I/\diff V$ values with a contour plot on the
three--dimensional tip trajectory, which is then projected on the two
dimensional plane of view, which in this paper will always be parallel to the
metal surface. The grayscale code associates bright areas to high conductance,
dark areas to low conductance.

The structure of the HOMO, LUMO and LUMO+1 molecular orbitals is clearly
visible in the calculated maps reported in~\fref{fig:C60bond66STSmaps},
similarly to what was observed in experiments (cf.\ Fig.\,3 of
\textcite{Lu2003} and Fig.\,3 of \textcite{Lu2004}).  In particular, the LUMO
level appears to be composed of two orbitals, LUMO$\alpha$ and LUMO$\beta$,
having a complementary spatial distribution of the conductance.  This pattern
was observed in experiments~\cite{Lu2003} for the Ag(100) surface, where the
LUMO peak displays a marked splitting of 0.4\,eV.  By calculating maps for two
energies (0.40 and 0.64\,eV) within the LUMO peak, we are able to extract two
conjugate LUMO images also for the case of the Au substrate, which points out
that the two states simply arise from the lifting of the degeneracy of the
LUMO level of the free molecule due to the contact with the surface.  We
therefore conclude that the splitting was not observable in the case of the Au
surface~\cite{Lu2004} due to the limits in resolution for STM spectroscopy.
What remains still unclear is the reason why the states are so well separated
in the experiments performed with the Ag(100) substrate~\cite{Lu2003}.  The
main clue is the large charge transfer to the molecule, compared to the
neutrality for the case of the Au surface, which may induce a deformation in
the molecule therefore lifting degeneracy of the LUMO level~\cite{Lu2003}.

\subsubsection{Abormal maps}

In order to investigate the sensitivity of the energy resolved maps to the
structure of the probe, we considered a truncated tip, having three atoms at
its end (\fref{fig:setup}c).  \Fref{fig:abnormal_images} shows the calculated
images for \Csixty\ with two different orientations of the tip (columns a and
b), whose position is given by the center of mass of the three end atoms.  At
all energies, the overall shape of the maps reflects the triangular
arrangement of the three atoms at the end of the truncated tip; a rotation of
the probe around the axis normal to the surface rotates the image of the
molecule (compare columns a and b of~\fref{fig:abnormal_images}), despite the
fact that the position of the molecule remains unchanged.  For the LUMO$+1$
level we observe that the bright rings corresponding to the pentagons in the
\Csixty\ are duplicated, which should be linked with the presence of two
adjacent atoms in the triangle.  For the orientation a) the alignment of the
tip is such that this sort of dichroism is present only for the left side of
the maps, while when the tip moves to the right side only one atom remains
closer to the molecule, and the single bright ring is recovered. A similar
behavior can be noticed also for LUMO$\alpha$ and the complementary map
LUMO$\beta$.

Most features described so far can be roughly explained assuming that, like in
the case of the sharp tip, at every position of the simulated scan tunneling
occurs between the sample and only one atom of the probe, namely the one
closest to the adsorbed molecule. The main difference is that for the
truncated tip such atom is not always the same, which leads to the dichroism
mentioned above. Only in the central region all three atoms may contribute
equally; the bright spot appearing right at the center of the LUMO$+1$ map
in~\fref{fig:abnormal_images} appears precisely because of the rise in
conductance given by the sum of the tunneling currents through the three end
atoms.  We also note that, compared to the well contrasted images
of~\fref{fig:topography_spectra}, the coherent structure of the HOMO level
becomes less evident when imaging with the blunt tip.  In this case, the
coarser nature of the probe sensibly changes the trajectory of the tip, which
therefore moves over a surface having a lower corrugation, borrowing form the
terminology used for flat surfaces. The tunneling current involves more than
one atom of the tip, hindering the resolution of the fine structure of the
HOMO orbital.

These results clearly demonstrate how the details of the energy resolved maps
of \Csixty\ can reveal the structure and orientation of the STM tip, and how
this kind of analysis could serve as a tool to characterize the properties of
the probes used in experiments.

\begin{figure}[t]
\begin{center}
\includegraphics[width=0.4\textwidth]{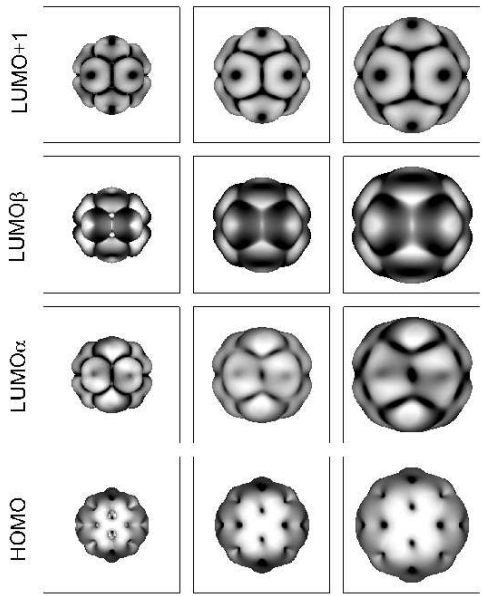}
\end{center}
\caption{Energy resolved maps from the LDOS for \Csixty\ on
  Ag(111) with a 6:6 bond on top. 
  The three columns from left to right show simulated tip
  trajectories which are approximately 2, 3.5, and 5\,\AA\
  away from the \Csixty. 
  The size of each frame is $20\times20$\,\AA.}
\label{fig:ldos_maps}
\end{figure}

\subsection{STM/STS imaging from LDOS}
\label{sec:LDOS}

Within Tersoff--Hamann theory\cite{Tersoff1985}, the STM topographic and
spectral images are modeled assuming that the current at bias $V$ is
proportional to the spatially resolved density of states:
\begin{equation}
I(\mathbf{r},V)\propto\int_{\efermi}^{\efermi+|e|V} \mathrm{d}E\;
\rho(\mathbf{r},E).
\label{eq:current_LDOS}
\end{equation}
Equation~(\ref{eq:current_LDOS}) allows to calculate the tip trajectory, based
on the assumption that the convolution effects due to the tip shape and states
can be neglected.  The local differential conductance $\diff
I(\mathbf{r},V)/\diff V\propto\rho(\mathbf{r},E)$ can be projected onto the
iso--current surface to simulate the experimental procedure followed to
measure STS maps.  The frames in~\fref{fig:ldos_maps} show that the patterns
in the molecular orbitals discussed so far can be reproduced with the LDOS,
provided that the calculated tip trajectory is placed far enough from the
molecule.  The standard choice is in fact to consider the LDOS at a few \AA\
from the atoms of the sample, and we verified that the peculiar LUMO$\alpha$
and LUMO$\beta$ images can be obtained only considering iso--surfaces at
3.5\,--\,5\,\AA\ from the \Csixty.

\section{Summary}

We have presented a detailed numerical investigation of STM topographic and
energy--resolved mappings of \Csixty\ on gold and silver surfaces.  The
transport calculations are based on non-equilibrium Green's function methods,
with a DFT--based tight--binding model. Self--consistency allows to study
charge redistribution on the molecule due to the presence of the supporting
surface, as well as due to external field generated by the voltage difference
between the substrate and the STM probe.

We have discussed the importance of realistic models for the tip; using a
sharp tip we could perfectly reproduce recently measured energy resolved maps
for \Csixty\ on metal surfaces\cite{Lu2003,Lu2004}.  The images are obtained
by projecting the contour plot of the differential conductance mapped on the
three--dimensional surface described by the tip as it scans the sample with a
fixed tunneling current.  Images of a similar quality can be obtained from the
LDOS using Tersoff--Hamann theory to calculate the conductance if the
simulated constant--current surface is sufficiently far from the cluster (here
more than 3\,\AA).

Additionally, we have modeled the STM/STS mapping with a truncated tip, and
described the appearance of abnormal images that allow to reconstruct the
structure of the probe by direct comparison with the ideal maps. We finally
suggest that this procedure could be used to gauge the properties of the STM
probes, which is generally unknown in STM experiments.

\section*{Acknowledgments}
We acknowledge financial support by the Deutsche Forschungsgemeinschaft (DFG)
through the priority program SPP 1153 ``Clusters in contact with surfaces:
Electronic structure and magnetism''.


\end{document}